\documentclass[10pt,conference]{IEEEtran}

\usepackage[T1]{fontenc}
\usepackage[utf8]{inputenc}
\usepackage{graphicx}
\usepackage{amsmath,amssymb}
\usepackage{booktabs}
\usepackage{balance}
\usepackage{url}

\graphicspath{{figures/}}

\newcommand{\hr}{\textsc{hashrope}}
\newcommand{\ms}{\,ms}
\newcommand{\us}{\,$\mu$s}

\begin{document}

\title{Hashrope: A Persistent Data Structure for Mechanisms of LLM Context
Management}

\author{\IEEEauthorblockN{Muntaser Syed, Ameerah Alsulami, and Marius Silaghi}
\IEEEauthorblockA{Florida Institute of Technology}}

\maketitle

\begin{abstract}
Serving stacks for large language models manage context through four
recurring mechanisms: incremental edits to a growing context, branching
and snapshotting for tree search, identification of reusable prefixes for
KV-cache reuse, and encoding of repeated segments. Each mechanism is
served today by a different specialist structure. We show that all four
reduce to a single persistent weight-balanced rope whose every node
carries a polynomial hash of its subtree over a Mersenne prime, so that
the structure supports logarithmic edits and logarithmic verified
content-identity queries at the same time. We validate each mechanism
with pre-registered, replicated experiments (nine runs per point,
deterministic operation-count guards), then compare against a deployed
specialist per mechanism: the Ropey editing library, a faithful
PagedAttention copy-on-write block table, and the vendored SGLang radix
matcher. On workloads that interleave edits with identity queries, \hr{}
is net-superior to Ropey by a factor that grows without bound (369.7x at
one million characters, 9/9 paired). Against the radix matcher on a live
vLLM stack serving Qwen2.5-7B-Instruct-1M on four A100 GPUs, \hr{}
identifies a one-million-token shared prefix exactly (0 mismatches in
1800 checks, non-vacuity established by 105/105 corrupted-entry
controls) and 1.20x faster (9/9 paired, p = 0.002), at a context length
the dense engine cannot even prefill. The same experiment measures what
prefix reuse is worth on real hardware: the engine's own prefix cache
reduces GPU energy per generated token by 69.6x to 277.6x as the shared
prefix grows, a saving we attribute to the engine, with the identity
layer adding exactly zero GPU work. The structure makes verified reuse a
first-class logarithmic query instead of a linear scan.
\end{abstract}


\section{Introduction}\label{sec:intro}

A language model serving stack spends much of its life doing string
bookkeeping. A conversation grows by appended turns and tool outputs;
an agent edits the middle of its working context; tree search forks one
context into many branches~\cite{yao2023tot}; schedulers look for
prefixes that two requests share so the KV cache of one can serve the
other~\cite{kwon2023pagedattention,zheng2024sglang,gim2024promptcache};
and system prompts, few-shot exemplars, and template headers repeat
across thousands of requests. These are four distinct mechanisms:
incremental edit, branch and snapshot, prefix-reuse identification, and
repetition encoding. In deployed systems each mechanism has its own
specialist. Text ropes handle edits~\cite{boehm1995ropes,ropey}.
PagedAttention shares KV blocks between branches through
copy-on-write block tables~\cite{kwon2023pagedattention}. SGLang finds
reusable prefixes with a radix tree over cached token
sequences~\cite{zheng2024sglang}. Repetition is handled ad hoc, usually
by materializing the bytes.

This paper makes a unification claim: all four mechanisms are served by
one structure. \hr{} is a persistent weight-balanced binary
rope~\cite{boehm1995ropes,nievergelt1973bounded} in which every node
additionally stores a polynomial fingerprint~\cite{karp1987efficient} of
the byte string its subtree represents, computed over a Mersenne prime
and maintained through every operation, in the spirit of a hash
tree~\cite{merkle1987digital} but over an editable, rebalancing
structure. Persistence~\cite{driscoll1989persistent} makes forks cheap;
the balance invariant makes edits logarithmic; and the maintained
fingerprints make content identity a logarithmic query rather than a
linear recomputation. A plain rope edits
in $O(\log N)$ but answers ``do these two contexts share a prefix, and
how long is it'' only by scanning. A radix cache answers the reuse
question but is not an editable representation of the context itself. A
hash tree certifies content but is not built for splits, joins, and
rebalancing. \hr{} does all of it in one object, so the context an
agent edits, the snapshot a search forks, and the key a scheduler
matches are the same thing.

The setting where the four mechanisms stop being four separate
problems is the long-lived agent. A coding or research agent over an
extended session holds one persistent context and subjects it,
interleaved, to all four operations: it edits the middle of the
buffer, forks it to explore alternatives, needs to know whether a
region is identical to something already processed downstream so a KV
cache or deduplicator can reuse it, and carries repeated boilerplate
across turns. Four specialists mean four representations of the same
object and conversions between them; one structure means the
operations compose. Whether deployed agent sessions actually occupy
the winning regime is a motivating hypothesis, not a claim we test;
what this paper establishes precisely is where that regime begins,
through per-mechanism crossovers and a query-rate ablation, so the
hypothesis is checkable.

We support the claim with three layers of evidence, all pre-registered
and replicated (nine runs per point, exact paired sign tests; protocol
in Section~\ref{sec:mechanisms}):

\begin{itemize}
\item \textbf{Mechanism validation} (Section~\ref{sec:mechanisms}).
Each mechanism is confirmed by a deterministic operation-count guard
and wall-clock scaling on real corpus data: exact tokenizer-aligned
ingestion, linear flatten with zero re-hashing,
$O(B\log N)$ branching memory, $O(\log^2 N)$ longest-common-prefix
queries, and $O(\log q)$ repetition encoding.
\item \textbf{Competitive studies} (Section~\ref{sec:competitive}).
One deployed specialist per mechanism: Ropey 1.6.1~\cite{ropey} for
edits (Rust against Rust), a faithful reimplementation of the
PagedAttention block-table copy-on-write
design~\cite{kwon2023pagedattention} for branching, and the vendored
SGLang RadixCache~\cite{zheng2024sglang} for identification. Wins and
losses are both reported; the losses delimit the regime where the
unification pays.
\item \textbf{Live serving-stack validation}
(Section~\ref{sec:realstack}). On vLLM with
Qwen2.5-7B-Instruct-1M~\cite{yang2025qwen1m} across four A100 GPUs,
\hr{} identifies shared prefixes exactly and, beyond a crossover the CPU
study predicted in advance, faster than the production radix matcher,
including at a one-million-token length the dense engine cannot serve.
The same runs measure what prefix reuse is worth in GPU energy,
attributed to the engine's cache.
\end{itemize}

Two results preview the flavor. First, on edit-plus-identity workloads
the advantage over the editing specialist grows without bound with
context size, reaching 369.7x at one million characters and 2901.8x at
ten million, because the fingerprint is read in $O(1)$ where the
baseline must recompute it in $O(N)$. That magnitude follows from the
cost model, since the specialist keeps no fingerprint; what the
experiment measures is the bounded price of maintaining one on raw
edits and the crossover it buys, which moves outward as predicted when
identity queries grow rarer. Second, in the stricter test, against a
radix matcher built for exactly this query, the live-stack crossover
falls between 524{,}288 and 1{,}000{,}000 tokens, and the earlier
CPU-only prediction, roughly 571k, lands inside that bracket: a
pre-registered quantitative prediction confirmed on real serving
hardware.

\section{The Hashrope Structure}\label{sec:structure}

\hr{} represents a byte string as an immutable binary tree with three
node kinds. A \emph{leaf} stores a block of bytes. An \emph{internal}
node represents the concatenation of its two children. A \emph{repeat}
node represents $q$ consecutive copies of its child. Every node stores
the length of the string it represents, its weight (leaf count), and a
polynomial hash of that string.

\textbf{Hashing.} For a byte string $S = s_0 s_1 \ldots s_{n-1}$ the
fingerprint is
$H(S) = \sum_{i=0}^{n-1} (s_i + 1)\, x^{\,n-1-i} \bmod p$
with $p = 2^{61}-1$ by default and base $x = 131$; the $+1$ offset
prevents leading-zero collisions, and Mersenne arithmetic keeps the
modular reduction to shifts and adds. Hashes compose: $H(A \Vert B) =
H(A)\, x^{|B|} + H(B) \bmod p$, so an internal node computes its hash
from its children in $O(1)$, and a repeat node computes
$H(S^q)$ through a geometric accumulator
$\Phi(q,\alpha) = \sum_{i=0}^{q-1}\alpha^i \bmod p$ evaluated by
repeated doubling in $O(\log q)$ multiplications. Every constructor
maintains these hash invariants, so the root hash of any \hr{} is the
fingerprint of the whole string at all times, and the fingerprint of a
freshly edited context costs one field read.

\textbf{Balance and edits.} The tree is kept weight-balanced in the
bounded-balance family~\cite{nievergelt1973bounded} using an Adams-style
balanced join~\cite{adams1993efficient}, which bounds height at
$O(\log w)$ for $w$ leaves. Concatenation is a balanced join; splitting
at a byte position descends one root-to-leaf path and rebuilds it,
touching $O(\log w)$ nodes; an insert is one split and two joins, a
delete two splits and one join. Because nodes are immutable, every
operation returns a new root that shares all untouched subtrees with
its input~\cite{driscoll1989persistent}: an edit allocates $O(\log w)$
nodes, and a fork allocates zero until the branches diverge.

\textbf{Substring hashes and prefix queries.} The fingerprint of any
substring is assembled from stored node hashes without materializing
bytes: a range decomposes into $O(\log w)$ fully covered subtrees,
whose stored hashes are combined by the composition rule, plus at most
two partial leaves. The longest common prefix of two ropes is then
found by binary search on prefix hashes, comparing two substring
fingerprints per step: exactly $2\lceil\log_2 N\rceil$ hash comparisons,
independent of content, for an $O(\log^2 N)$ identity query that never
scans the text.

\textbf{Byte-index contract.} The structure stores, splits, and hashes
bytes; all indices are byte offsets. Splitting UTF-8 text at an
arbitrary byte offset is exact over bytes (the halves concatenate back
to the original and all hashes are over bytes), and character-boundary
alignment is the caller's concern; the tokenizer-aligned chunker of
Section~\ref{sec:mechanisms} is such a caller, and four-byte token words
(Section~\ref{sec:competitive}) align at every multiple of four. Python
and Rust implementations exist; the CPU studies use Python except the
Rust-against-Rust edit study. Both define the same function, since a
fingerprint depends only on the bytes and on $(p, x)$, never on tree
shape, with the same defaults and offset; no cross-implementation
conformance run exists, so agreement rests on the shared definition,
not on measurement (Rust's lazy mode holds a sentinel until a hash is
materialized).

\section{Four Mechanisms, One Structure}\label{sec:mechanisms}

We first confirm that each mechanism behaves as the analysis says it
must, on real corpus data. Every experiment follows one protocol:
pre-registered promotion criteria written before any data; three corpus
seeds crossed with three independent process invocations (nine runs per
cell); mean and standard deviation reported across runs; and, wherever
the claim is structural, a deterministic operation-count guard whose
value is a proof rather than an estimate. The error model was itself
validated: an independent repeat of an early pilot showed within-run
bootstrap intervals understate run-to-run variance (process spawn,
scheduling, thermal state), so within-run intervals were retired and
the spread across independent invocations is the reported uncertainty
throughout. Absolute Python timings carry
an interpreter constant; for those results the transferable content is
the guard-proven operation count and the measured scaling exponent, and
the Rust edit study in Section~\ref{sec:competitive} shows the direct
numbers. Table~\ref{tab:mechanisms} summarizes.

\begin{table}[t]
\caption{Mechanism validation. Guards are deterministic (std = 0);
timings are mean over 9 runs. Slopes are log-log wall-clock exponents.}
\label{tab:mechanisms}
\centering
\footnotesize
\begin{tabular}{@{}llll@{}}
\toprule
Mechanism & Bound & Guard & Headline \\
\midrule
Ingestion & exact tokens & 0 divergences & 7.6x at 16-way \\
Flatten & $O(N)$ & 0 re-hash / re-alloc & 800x vs re-split \\
Branching & $O(B\log N)$ mem & $O(\log w)$/fork & 166x at 8\,MB \\
Identify & $O(\log^2 N)$ & $2\lceil\log_2 N\rceil$ steps & slope 0.028 \\
Repetition & $O(\log q)$ & 1 node per repeat & $10^6$x at $q{=}10^4$ \\
\bottomrule
\end{tabular}
\end{table}

\textbf{Ingestion with exact token identity.} Snapping leaf boundaries
to tokenizer-safe split points lets disjoint leaves be tokenized in
parallel while reproducing the whole-string token IDs exactly: zero
boundary divergences across byte-BPE, SentencePiece, and WordPiece
families. The fixed-byte control fails that guarantee in both ways that
matter: its token stream first diverges near token 1800 (a 0.085 percent
token-count drift on prose), and its blind cuts bisect multibyte
codepoints, leaving six invalid UTF-8 leaves; the aligned chunker
produces zero of either, through a single boundary predicate verified by
a global gate. Given the identity, chunking alone yields a stable single-thread gain
(1.46x to 1.50x for GPT-2, 1.30x to 1.36x for T5, std at most 0.02)
at 1\,MB to 4\,MB contexts, and
16-way parallelism reaches 6.62 $\pm$ 0.37 (thread pool) and 7.59
$\pm$ 0.41 (multiprocessing) on GPT-2 tokenization of 4\,MB inputs,
with multiprocessing beating threads 9/9 (p = 0.004) at a modest 13 to
16 percent margin. One exception: at 1\,MB with the T5 tokenizer,
threads edge multiprocessing (not significant). Contexts up to
256\,KB are overhead-dominated and excluded.

\textbf{Flatten in linear time.} Materializing the rope to contiguous
bytes by in-order leaf traversal performs zero splits, zero leaf
re-allocations, and zero hash recomputations (guard), completing 2\,MB
in 0.546 $\pm$ 0.016\ms{} against 437 $\pm$ 60\ms{} for a midpoint
re-splitting strategy that re-hashes every byte, a roughly flat 730x to
800x advantage for inputs of 1\,MB and above. Both paths scale
linearly (log-log slopes 1.034 and 1.044); the eliminated cost is
redundant re-hashing, not tree depth. Table~\ref{tab:flatten} gives
the sweep with its guard counts; sizes are decimal rather than powers
of two, since power-of-two leaf counts land every midpoint on a leaf
boundary and would hide the control arm's cost.

\begin{table}[t]
\caption{Flatten sweep, mean over nine runs. The guard column counts
leaf re-allocations / splits / hash recomputations per flatten; the
in-order materializer performs none at any size.}
\label{tab:flatten}
\centering
\footnotesize
\begin{tabular}{@{}rrrrr@{}}
\toprule
$N$ (B) & re-split (ms) & in-order (ms) & guard L/S/H & speedup \\
\midrule
64\,K  & 11.2  & 0.003 & 30/15/30 & 3427x \\
256\,K & 48.9  & 0.050 & 126/63/126 & 1082x \\
1\,M   & 199.2 & 0.272 & 510/255/510 & 735x \\
2\,M   & 437.4 & 0.546 & 1022/511/1022 & 800x \\
4\,M   & 865.0 & 1.166 & 2044/1023/2044 & 741x \\
8\,M   & 1692.6 & 2.305 & 4092/2047/4092 & 734x \\
\bottomrule
\end{tabular}
\end{table}

\textbf{Branch and snapshot in $O(B\log N)$ memory.} Forking a context
allocates only the root-to-leaf spine when a branch first diverges. The
node-count guard is exactly deterministic: at $N = 2$\,M and $B = 50$
branches, the rope arms hold 1{,}477 unique objects against 49{,}927
for deep copies (33.8x structural sharing), with about
$\lceil\log_2 w\rceil$ new nodes per fork, matching the bound. Measured
memory compression grows with context size, reaching 166x at $N = 8$\,M
with $B = 100$ (159\,KB against 25.8\,MB), and per-fork rope memory is
near constant (slope 0.183) while deep-copy memory is linear (slope
0.884). The compression is stable in the number of branches, 50.9x at
fifty against 50.2x at one hundred at 2\,M, because both arms are
linear in branch count, a fact the paged-baseline branch-count
ablation in Section~\ref{sec:competitive} confirms independently.
Fork creation takes about 0.037\ms{} independent of $N$,
against 4 to 17\ms{} for deep copies (114x to 454x). The bound is
$O(B\log N)$, correcting an earlier $O(1)$ intuition, and the data
follow it.

\begin{figure}[t]
\centering
\includegraphics[width=\columnwidth]{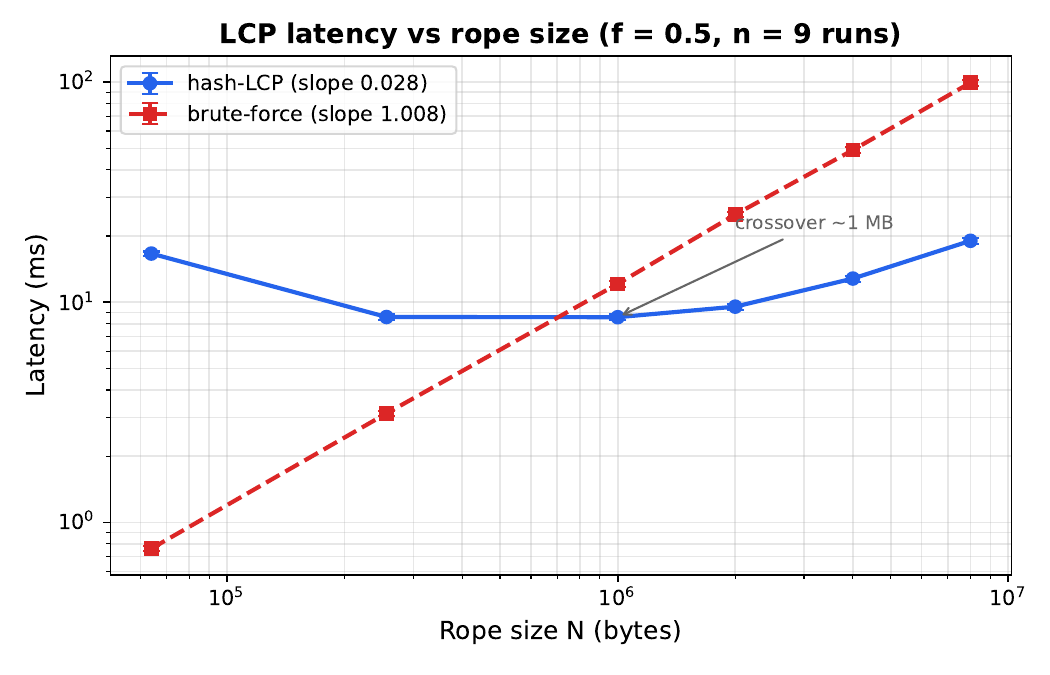}
\caption{Prefix-identification latency over context size at half-way
divergence, nine runs per point, log-log. The hash LCP is near flat
(the $O(\log^2 N)$ signature) while the byte scan is linear; the
crossover falls near 1\,MB, where long-context serving begins.}
\label{fig:lcp}
\end{figure}

\textbf{Prefix-reuse identification in $O(\log^2 N)$.} The
binary-search LCP performs exactly $2\lceil\log_2 N\rceil$ hash
comparisons at every size (guard), and wall clock corroborates the
bound with a log-log slope of 0.028 across 64\,KB to 8\,MB, against
1.008 for byte-by-byte comparison; the crossover with the byte scan
falls near 1\,MB. At 8\,MB the hash LCP completes in 19\ms{} wherever
the divergence lies, against 99 to 200\ms{} for scanning. Correctness
is exact: zero mismatches against a byte-level oracle in 162 checks.
One boundary: when two contexts diverge at the very first byte,
the scan exits immediately while the binary search still pays its
logarithmic steps, so trivially disjoint contexts belong to the scan.
A dedup workload with ten prompts sharing a 1.2\,MB prefix resolves
all 45 pairs at 20.9\ms{} per pair; in 44 the query returns the
constructed prefix exactly, and in the other a longer one, because the
two suffixes continue identically past the constructed boundary, as
the byte oracle confirms. Fig.~\ref{fig:lcp} shows the scaling
contrast.

\textbf{Repetition in $O(\log q)$.} Encoding $q$ copies of a unit
creates one repeat node wrapping the unit subtree and computes the hash
through the geometric accumulator. At $q = 10^4$ with a 4\,KB unit,
that is 2 objects, 327 bytes, and 11\us{} to represent 40\,MB of
logical content, against 19{,}999 nodes, 44\,MB, and 10.6\,s for naive
materialization: a node compression near $10^4$ and a construction
speedup near $10^6$, with construction slopes 0.175 (repeat) against
1.041 (naive). A unit-size sweep separates the two costs: naive
construction grows linearly with unit length, while the repeat-node
cost stays near 0.01\ms{} regardless of the unit. Repeated system
prompts, few-shot exemplars, and template headers are encoded once.

Table~\ref{tab:ledger} aggregates the verification machinery of every
study: each pairs an independent correctness oracle with a
deterministic structural guard and, wherever exactness is claimed, an
injected fault proving the gate can fail. The mismatch column is zero
everywhere; the controls column is why those zeros mean something.

\begin{table}[t]
\caption{Verification ledger across all studies. Every oracle
comparison in the paper returned zero mismatches; negative controls
(injected faults) confirm each gate is non-vacuous.}
\label{tab:ledger}
\centering
\footnotesize
\begin{tabular}{@{}p{1.55cm}p{2.6cm}p{2.9cm}@{}}
\toprule
Study & Oracle / gate (0 mismatches) & Guard and negative control \\
\midrule
Ingestion & whole-string token IDs, 3 tokenizer families & fixed-byte control fails: drift at token 1800, 6 invalid UTF-8 leaves \\
Flatten & bytes equal original and control arm, all cells & op counts: 0 splits, 0 re-allocs, 0 re-hashes vs $\Theta$(leaves) \\
Identify & byte-level scan, 162 checks & steps $= 2\lceil\log_2 N\rceil$ exactly, content-independent \\
Branch & fork materialization, all cells & node count deterministic (std $=$ 0), $O(\log w)$ per fork \\
Repeat & bytes and hash equal naive arm & node count $=$ unit $+$ 1 exactly \\
vs Ropey & byte-identity and maintained hash $\equiv$ recompute, every edit step & bit-flip on maintained hash fails the gate \\
vs Paged & token oracle, every node, both milestones & spine $\le \lceil\log_2 w\rceil{+}3$; fork entries $= \lceil N/B\rceil$ \\
vs radix & token oracle, all cells & instrumented matcher proves $\ge L$ comparisons; \hr{} $\le 2\lceil\log_2 N\rceil$ \\
Live stack & three arms token-identical, 1800 checks & 105/105 corrupted entries detected at exact position \\
\bottomrule
\end{tabular}
\end{table}

\section{Against the Specialists}\label{sec:competitive}

Unification is only interesting if the unified structure is competitive
with each specialist on its home ground and clearly better once
identity enters the workload. We test one deployed implementation per
mechanism; setup is never timed, and all comparisons are paired on
identical seeded workloads.

\begin{figure}[t]
\centering
\includegraphics[width=\columnwidth]{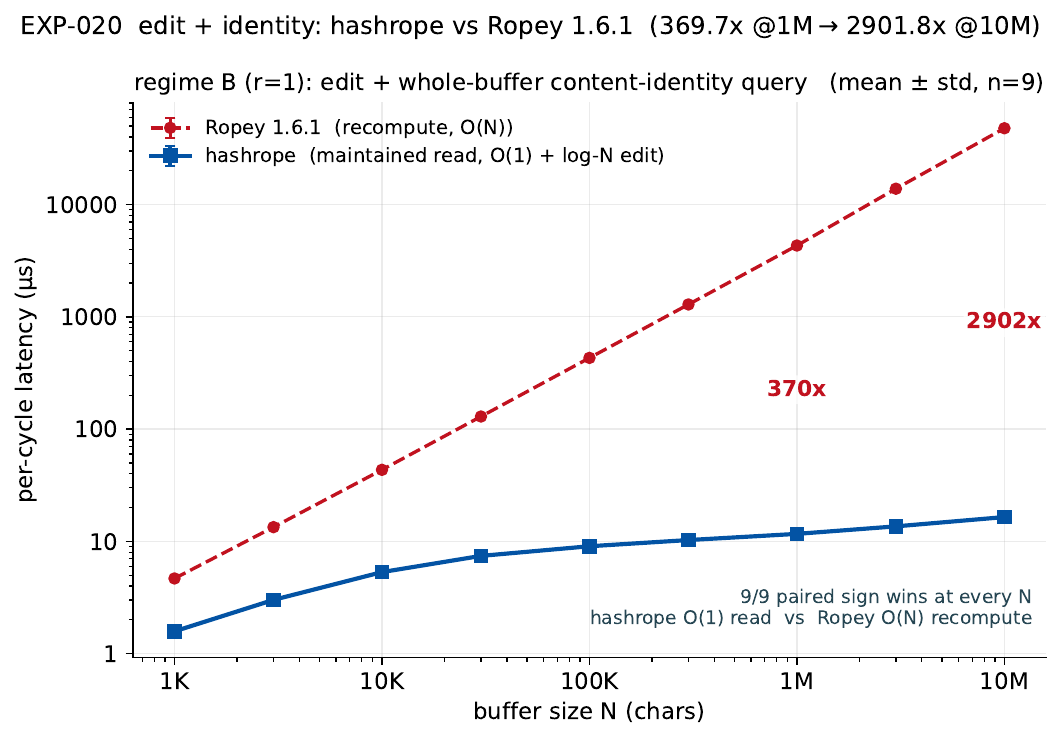}
\caption{Edit-plus-identity workload, \hr{} (Rust) against Ropey 1.6.1
with the same polynomial hash: per-cycle latency over buffer size, nine
runs per point. Ropey's cycle is dominated by the $O(N)$ fingerprint
recompute (slope 1.013); \hr{} reads the maintained fingerprint in
$O(1)$ (slope 0.150), so the gap widens without bound, 9/9 paired at
every size.}
\label{fig:editidentity}
\end{figure}

\textbf{Edits: against Ropey (Rust against Rust).} Ropey
1.6.1~\cite{ropey} is a production text rope used by editors. We vendor
it (pinned, checksum verified) and drive both arms with one seeded
churn sequence (delete a random span, insert an equal-length span) on
ASCII buffers so character and byte indices coincide. \hr{} maintains
its fingerprint eagerly on every edit; Ropey, which maintains no hash,
is given \hr{}'s own polynomial hash to run over its materialized bytes
when identity is queried, so the comparison is maintained-incrementally
against recomputed-from-scratch with the same hash function.
Correctness is gated as in every study here: the two arms stay
byte-identical, and the maintained fingerprint equals a from-scratch
recompute, at every edit step of every run with zero mismatches, made
non-vacuous by a bit-flip negative control on the maintained hash.
One configuration choice conservatively handicaps \hr{}: its leaf cap
of 512 bytes is finer than Ropey's roughly 1\,KB chunk target, giving
\hr{} the deeper tree. Two findings. On raw churn, both arms are in the same $O(\log N)$ class
(per-edit slopes 0.202 against 0.212); \hr{} pays a bounded constant
factor for the maintained fingerprint that peaks at 22.0x near 100\,K
and declines to 14.0x at 10\,M as both arms go logarithmic. On the
workload that motivates a content-indexed structure, cycles of edits
followed by a whole-buffer identity query, \hr{} wins at every size and
the advantage grows without bound: 3.0x at 1\,K, 47.6x at 100\,K,
369.7x at 1\,M (9/9 paired, p = 0.0039), 2901.8x at 10\,M
(Fig.~\ref{fig:editidentity}). The mechanism is visible in the
per-cycle slopes: 1.013 for Ropey (the $O(N)$ recompute dominates)
against 0.150 for \hr{} (an $O(1)$ fingerprint read plus logarithmic
edits). Because that argument depends on how often identity is
queried, we ablate the query-to-edit ratio $r$
(Table~\ref{tab:rsweep}). The crossover moves outward exactly as the
cost model predicts, from 1\,K characters at one query per cycle to
300\,K at one query per hundred cycles, and below each crossover Ropey
wins (down to 0.2x at small $N$ under $r = 0.01$), which delimits the
regime. Above the crossover \hr{} wins 9/9 at every
size, and even at $r = 0.01$ the advantage reaches 27.7x at 10\,M,
because Ropey's per-cycle slope stays near one at every ratio (1.013,
1.007, 0.943) while \hr{}'s stays near 0.2. One more cost: persistence
retains dead path nodes under in-place churn, about 2.7 per doubling
of $N$, the flip side of the immutability that makes forks free.

\begin{table}[t]
\caption{Query-to-edit ratio ablation against Ropey: per-cycle speedup
(Ropey over \hr{}), nine runs per point, 9/9 paired at and above each
crossover $N^{*}$. Below $N^{*}$ Ropey wins.}
\label{tab:rsweep}
\centering
\footnotesize
\begin{tabular}{@{}lrrrr@{}}
\toprule
$r$ (queries/cycle) & $N^{*}$ & 100\,K & 1\,M & 10\,M \\
\midrule
1    & 1\,K   & 47.6x & 369.7x & 2901.8x \\
0.1  & 10\,K  & 5.3x  & 32.2x  & 265.0x \\
0.01 & 300\,K & 0.7x  & 3.8x   & 27.7x \\
\bottomrule
\end{tabular}
\end{table}

\textbf{Branching: against PagedAttention copy-on-write.} We
reimplement the block-table copy-on-write design of
PagedAttention~\cite{kwon2023pagedattention} faithfully (ref-counted
physical blocks, per-sequence block tables, copy-on-write fork and
append; 55 unit tests). Three latency decompositions are measured and
all three reported. On a bare fork, \hr{} shares the immutable root in
about 39\,ns, flat at every $N$, while the baseline copies its block
table (1.1\us{} to 6.13\ms{}): \hr{} wins everywhere, but only
because divergence is deferred, so the gated metric was locked before
any data to branch creation, fork plus the first divergent step, which
charges \hr{} its full $O(\log w)$ spine and admits a genuine
crossover (Table~\ref{tab:branchlat}). On per-token append the specialist wins, 0.33\us{} to
1.0\us{} amortized $O(1)$ against \hr{}'s 11\us{} to 17\us{}
$O(\log w)$, an 18x to 35x advantage pre-registered as the bounded
price of generality.

\begin{table}[t]
\caption{Branch-creation latency against the paged copy-on-write
baseline over context size, reference block 16, nine runs per point;
all sizes 9/9 paired at 1\,M for every block size.}
\label{tab:branchlat}
\centering
\footnotesize
\begin{tabular}{@{}rrrrr@{}}
\toprule
$N$ (tok) & \hr{} (\textmu s) & paged b16 & speedup b16 & b8 \\
\midrule
1\,K   & 25.9 & 13.9\us{}  & 0.54x & 0.64x \\
4\,K   & 30.1 & 25.9\us{}  & 0.86x & 1.30x \\
16\,K  & 31.6 & 62.3\us{}  & 1.97x & 3.87x \\
64\,K  & 33.8 & 219.5\us{} & 6.49x & 11.6x \\
256\,K & 35.0 & 834.9\us{} & 23.9x & 44.9x \\
1\,M   & 37.1 & 3.61\ms{}  & 97.4x & 168.7x \\
\bottomrule
\end{tabular}
\end{table}

\begin{figure}[t]
\centering
\includegraphics[width=\columnwidth]{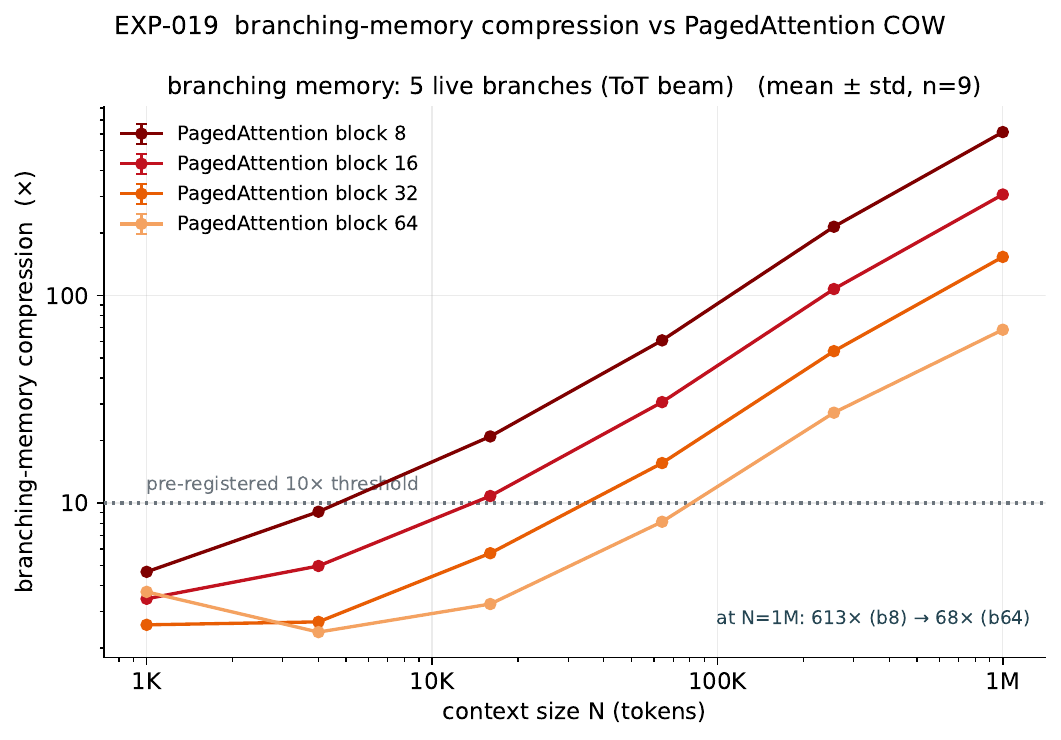}
\caption{Branch memory against a faithful PagedAttention block-table
copy-on-write baseline, five branches per context. The paged table is
$\Theta(N/B)$ per branch; \hr{} adds an $O(\log w)$ shared spine per
branch, so compression grows with context size.}
\label{fig:branchmem}
\end{figure}

\hr{} branch creation is essentially flat, 26\us{}
to 37\us{} from 1\,K to 1\,M tokens, while the block table scales as
$O(\lceil N/B\rceil)$, reaching 3.6\ms{} at 1\,M with block size 16:
crossover at 16\,K tokens for the reference block size, 97.4x at 1\,M
(168.7x at block size 8), 9/9 paired at every block size. Holding five
branches of a 1\,M-token context costs \hr{} about 9.2\,KB against
5.63\,MB for the paged table at block size 8 (613x; 306x at block 16),
because the table is $\Theta(N/B)$ per branch while \hr{} adds only the
shared logarithmic spine (Fig.~\ref{fig:branchmem}). Both sides carry
deterministic guards at every cell: \hr{} allocates at most
$\lceil\log_2 w\rceil + 3$ new nodes per branch, and the baseline's
fork copies exactly the block-table entries the design specifies. A
branch-count ablation over $B \in \{5, 10, 25, 50\}$ shows the
compression is invariant to beam width, rising slightly from 613x to
639x at block size 8 as $B$ grows (both sides scale linearly in $B$);
fifty branches of a 1\,M-token context fit in 86\,KB against
53.7\,MB. An ecological replay corroborates the controlled sweep with
real search: tree-of-thought traces~\cite{yao2023tot} generated by a
120B open-weight model on Game of 24 (75 trees, 7{,}943 nodes,
7{,}868 branch creations at beam 5, depth 3), recorded once and
replayed deterministically through the same driver over prompt
prefixes up to 256\,K tokens. Every node context reconstructs
byte-identically in both arms (0 mismatches at every prefix and block
size), \hr{} branch creation stays flat at 16\us{} to 24\us{} while
the paged table reaches 1.82\ms{}, giving 74.6x, 38.9x, 19.1x, and
9.6x at the 256\,K prefix for block sizes 8 through 64 (9/9 paired at
every block size). Marginal branching memory on the real
five-survivor beams compresses 67x, 34x, and 17x across the
deployment-standard block sizes 8 to 32, the whole beam fitting in
about 480\,KB against 4.6\,MB to 32\,MB; at the largest block, 64,
compression is 9.6x, just under the pre-registered 10x bar, the
sensitivity tail of the block-size trend. The crossover is
block-dependent: block 8 crosses by a 4\,K-token prefix, blocks 16 and
32 by 16\,K, block 64 by 64\,K, which tells an operator where the
tradeoff sits for their configuration. Below those crossovers the
paged table wins, as on the bare puzzles with no prefix (a few-token
base, 0.30x to 0.40x); Table~\ref{tab:mb} spans the whole regime.

\begin{table}[t]
\caption{Ecological replay of real 120B-model tree-of-thought search:
branch-creation speedup (paged over \hr{}) by prepended prefix length
$P$, nine runs per point; sub-crossover cells where the paged table
wins are included.}
\label{tab:mb}
\centering
\footnotesize
\begin{tabular}{@{}rrrrr@{}}
\toprule
$P$ (tok) & \hr{} (\textmu s) & paged b16 (\textmu s) & speedup b16 & b8 \\
\midrule
0      & 16.0 & 5.6  & 0.35x & 0.40x \\
4\,K   & 17.5 & 20.0 & 1.14x & 2.14x \\
16\,K  & 21.5 & 67.1 & 3.13x & 7.81x \\
64\,K  & 23.3 & 239  & 10.3x & 20.9x \\
256\,K & 24.4 & 948  & 38.9x & 74.6x \\
\bottomrule
\end{tabular}
\end{table}

\textbf{Identification: against the SGLang radix matcher.} We vendor
the RadixCache of SGLang~\cite{zheng2024sglang} byte-identically
(v0.1.17, checksum verified) and compare single-query prefix
identification over token sequences, with a NumPy flat scan as the
C-speed floor. Token streams enter \hr{} as fixed-width four-byte
words, a deliberately conservative encoding (two bytes would halve the
byte length and shrink the logarithm in \hr{}'s favor). The radix
walk's $\Theta(L)$ cost is not assumed but proven: a deterministically
generated instrumented copy of the vendored matcher, used only for
operation counting and never timed, confirms at least $L$ token
comparisons on every full-prefix query, while \hr{}'s step guard holds
at $2\lceil\log_2 N\rceil$ per query. The walk costs about 29\,ns per token;
the \hr{} LCP is near flat at 13\ms{} to 18\ms{}
(Table~\ref{tab:radix}). The crossover on CPU
falls near $L^* \approx 571$\,K tokens: 1.70x at 1\,M and 4.63x at
2\,M (9/9 paired, p = 0.004), with the advantage growing linearly in
$L$ beyond.

\begin{table}[t]
\caption{Single-query prefix identification against the vendored
SGLang radix matcher over shared-prefix length $L$ (tokens), mean over
nine runs; the flat scan is the $O(L)$ floor. Crossover near 571\,K.}
\label{tab:radix}
\centering
\footnotesize
\begin{tabular}{@{}rrrrr@{}}
\toprule
$L$ & radix (ms) & \hr{} (ms) & flat (ms) & speedup \\
\midrule
1\,K   & 0.034  & 17.83 & 0.003 & 0.00x \\
4\,K   & 0.118  & 17.77 & 0.003 & 0.01x \\
16\,K  & 0.462  & 14.32 & 0.006 & 0.03x \\
64\,K  & 2.24   & 13.07 & 0.018 & 0.17x \\
128\,K & 3.82   & 15.57 & 0.038 & 0.25x \\
256\,K & 7.47   & 15.53 & 0.082 & 0.48x \\
512\,K & 15.09  & 15.95 & 0.171 & 0.95x \\
1\,M   & 29.23  & 17.23 & 0.404 & 1.70x \\
2\,M   & 58.76  & 12.70 & 1.646 & 4.63x \\
\bottomrule
\end{tabular}
\end{table}

Two negatives delimit the regime. On real
conversation pairs from ShareGPT~\cite{sharegpt} and
LMSYS-Chat-1M~\cite{zheng2024lmsyschat}, whose shared prefixes average
about 2\,K tokens, the radix matcher wins every cell by orders of
magnitude (0.03\ms{} against 8.8\ms{}), and in a one-against-$K$ sweep
the radix tree answers all $K$ candidates in one walk while \hr{} pays
$K$ pairwise queries (610x to the radix side at $K = 100$). The
pairwise form belongs to the query, not the structure: whether any of
$K$ cached contexts shares a length-$\ell$ prefix with the query is
monotone in $\ell$, and any block-aligned prefix fingerprint is one
substring query (a chained block hash~\cite{vllm2025apc} must first
compute every earlier key), so an index of block-aligned prefix
fingerprints answers one-against-$K$ by one binary search of
$O(\log N)$ probes, independent of $K$, at $O(N/b)$ entries per
context for block size $b$. This is design analysis; the measured
negative stands. The identification advantage claimed here is for
long shared prefixes, one pair at a time. The pattern
across all three studies is the thesis in miniature: each specialist
buys its per-operation constant by giving something up elsewhere (the
flat scan's cheap query costs $O(N)$ edits, the radix tree's cheap
match costs $O(N)$ materialized state per cached context, Ropey's
cheap edit costs $O(N)$ per identity query), while \hr{} pays a
higher per-query constant once and amortizes it across all four
mechanisms simultaneously.

\section{Validation on a Live 7B Serving Stack}\label{sec:realstack}

The CPU studies leave open whether any of this survives a real serving
engine at lengths that matter, with correctness verified end to end and
energy measured rather than asserted. We answer on vLLM 0.8.5.post1
serving Qwen2.5-7B-Instruct-1M~\cite{yang2025qwen1m} with tensor
parallelism across four A100-40GB GPUs.

\textbf{Design.} Two decoupled tracks share every run. Track A varies
exactly one engine knob, the prefix cache (off, on), and measures GPU
energy through NVML on-board energy counters, time to first token, and
throughput; the long-context serving configuration (chunked prefill,
eager execution, single-sequence batching) is held constant across the
cache axis, so the cache flag is the only difference in dispatched
work. Track B runs the three identification arms (\hr{}, the
vendored radix matcher, the flat scan) on CPU, strictly outside the GPU
measurement window, over the same token streams the engine serves, so
the identity layer never touches the energy measurement: with one
engine execution per cell, the identifier arms differ by exactly zero
dispatched GPU work, and energy neutrality holds by construction. Two
regimes: realistic multi-turn
streams drawn from ShareGPT and LMSYS-Chat-1M (median shared prefixes
near 642 tokens), and a controlled sweep of a single shared prefix of
length $L$ in $\{$131{,}072; 262{,}144; 393{,}216; 524{,}288;
1{,}000{,}000$\}$ tokens followed by short divergent queries. Nine runs
per point as before.

\textbf{Exactness, made non-vacuous.} Across every cell and seed, the
reuse prefix selected by \hr{}, the radix matcher, and the flat scan is
token-identical: 0 mismatches in 1800 checks. A gate that trivially
agreed would also score zero, so we corrupt cached entries on purpose:
a single token flipped at the first, middle, or last position of the
cached prefix, 105 controls in all, including position 999{,}999 of a
1{,}000{,}000-token prefix. Every arm detects every corruption at the
exact position, 105/105. Verified reuse here means verified. The
realistic streams also reproduce the short-prefix negative on the live
stack: at their roughly 642-token median prefixes the radix matcher
identifies in 0.037 $\pm$ 0.002\ms{} against 17.46 $\pm$ 0.91\ms{} for
\hr{} (flat scan 0.005\ms{}), consistent with the CPU study and
reported in full. Two measurement notes for reproducibility. The
engine's offline API exposes no per-request cached-token counter in
this version, so the tie between identified prefixes and actual engine
reuse is carried by the time-to-first-token and energy deltas together
with the correctness gate, not by a counter. And energy per token
divides total GPU Joules by generated completion tokens, so absolute
values are prefill-dominated; the cache-off over cache-on ratio and
the paired delta are the quantities we report.

\begin{figure}[t]
\centering
\includegraphics[width=\columnwidth]{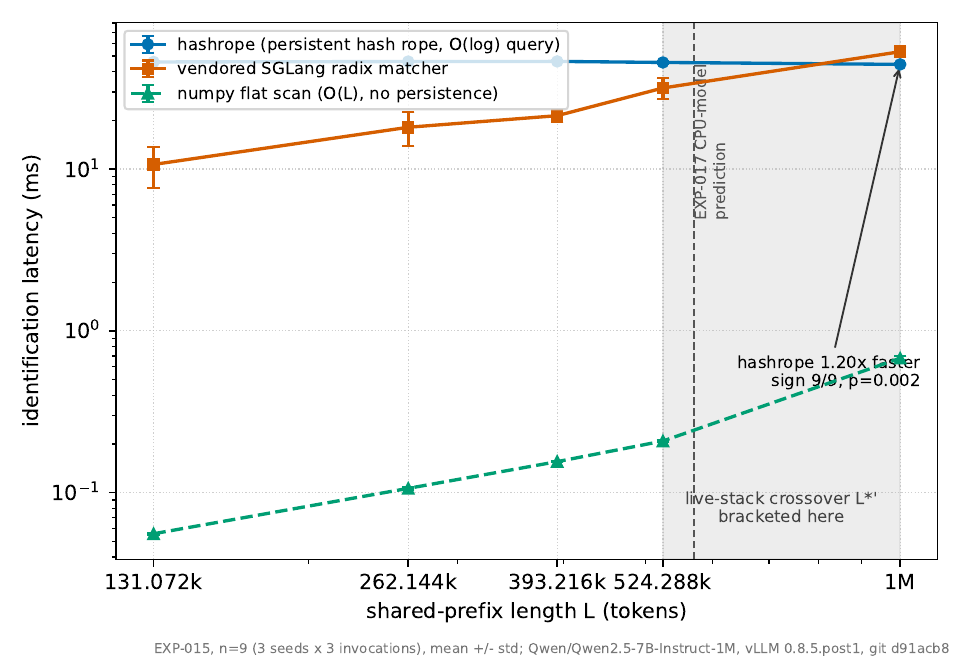}
\caption{Identification latency on the live stack over shared-prefix
length, log-log, nine runs per point. The live-stack crossover $L^{*\prime}$
is bracketed between 524{,}288 and 1{,}000{,}000 tokens; the dashed line
is the crossover predicted in advance by the CPU-only study, which
lands inside the bracket.}
\label{fig:crossover}
\end{figure}

\textbf{The crossover, confirmed on live hardware.} The CPU study of
Section~\ref{sec:competitive} predicted a crossover near 571\,K tokens
before any serving run. On the live stack the prediction holds:
the radix matcher is still faster at $L = 524{,}288$ (31.73 $\pm$
4.64\ms{} against 45.66 $\pm$ 0.76\ms{}) and \hr{} is faster at
$L = 1{,}000{,}000$ (44.42 $\pm$ 0.63\ms{} against 53.19 $\pm$
3.78\ms{}; 1.20x, 9/9 paired, one-sided p = 0.002), bracketing the
live-stack crossover inside the predicted region
(Fig.~\ref{fig:crossover}). \hr{}'s latency is flat across the entire
sweep, the logarithmic signature, while the radix walk grows linearly.
At one million tokens the dense-attention engine cannot complete a
prefill in this configuration: it initializes cleanly and allocates KV
cache for 1.39x
the full request across the four GPUs, but the first dense $O(L^2)$
prefill step never returns within the engine's execution window (the
runs record the wall explicitly and skip the engine track there), yet the identity layer keeps answering in
44\ms{}, which is 0.048 percent of the full prefill time at the
largest length the engine does serve. The serving wall belongs to the
engine configuration; the identification does not share it.

\begin{figure}[t]
\centering
\includegraphics[width=\columnwidth]{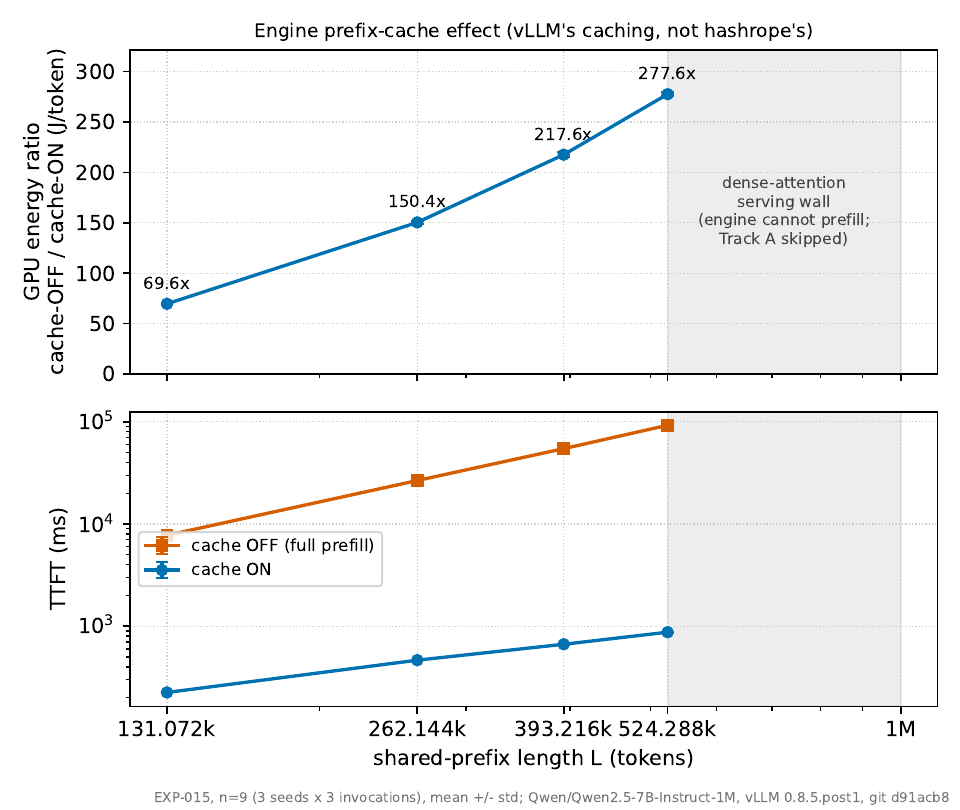}
\caption{The engine prefix-cache effect on the live stack (top: GPU
energy per generated token, cache off over cache on; bottom: time to
first token), nine runs per point. The reduction, 69.6x to 277.6x
across the sweep, is attributed to the vLLM cache, not to \hr{}. The shaded
region marks the dense-attention serving wall, beyond which the engine
cannot prefill; identification (Fig.~\ref{fig:crossover}) keeps
working there.}
\label{fig:energy}
\end{figure}

\textbf{What reuse is worth.} With the cache on,
the engine's own prefix caching reduces GPU energy per generated token
by 22.65x $\pm$ 1.84 on the realistic streams and by 69.6x, 150.4x,
217.6x, and 277.6x at $L$ from 131\,K to 524\,K on the controlled
sweep, monotone in $L$, 9/9 paired with p = 0.002 at every point; time
to first token falls from 92.5\,s to 0.87\,s at $L = 524{,}288$, with
zero cache-on regressions anywhere (Fig.~\ref{fig:energy}). Table~\ref{tab:energy} gives the exact values.

\begin{table}[t]
\caption{Live-stack prefix-cache effect, mean over nine runs, all
rows 9/9 paired with p $=$ 0.002. The reduction is the engine's; the
identity layer adds zero GPU work by construction.}
\label{tab:energy}
\centering
\footnotesize
\begin{tabular}{@{}lrrrr@{}}
\toprule
Cell & J/tok off & J/tok on & ratio & TTFT off $\to$ on \\
\midrule
Realistic & 2.48 & 0.110 & 22.7x & 56\ms{} $\to$ 21\ms{} \\
$L{=}131\,K$ & 1555 & 22.3 & 69.6x & 7.8\,s $\to$ 0.22\,s \\
$L{=}262\,K$ & 5258 & 35.0 & 150.4x & 26.6\,s $\to$ 0.47\,s \\
$L{=}393\,K$ & 10901 & 50.1 & 217.6x & 54.6\,s $\to$ 0.67\,s \\
$L{=}524\,K$ & 18131 & 65.3 & 277.6x & 92.5\,s $\to$ 0.87\,s \\
\bottomrule
\end{tabular}
\end{table} The
reduction is the effect of the engine's cache, not of \hr{}. What
\hr{} contributes is the verified identification that decides reuse,
at exactly zero additional GPU work and at millisecond latencies that
are a rounding error against prefill. An upstream identity layer is
redundant inside a single engine that already prefix-matches its own
cache (a radix tree in SGLang, chained block hashes in
vLLM~\cite{vllm2025apc}); the measured stakes say what verified
identification is worth in the settings where the engine's matcher does
not reach: routing across engine instances, cross-request
deduplication, and contexts held as opaque compressed blobs.

\textbf{What integration would take.} Replacing an engine's matcher
touches one hook, the function that turns a request into prefix keys:
SGLang walks its radix tree token by token, and vLLM hashes each full
block with its parent's key~\cite{vllm2025apc}. Keyed by \hr{} prefix
fingerprints, a maintained context supplies any block-aligned key as
one substring query, while the engine keeps allocation, reference
counting, and eviction. Inside one engine that saves matching work
already negligible against prefill, so the integration that pays sits
one layer up, in a router or deduplicator holding one fingerprint index
across instances. Admission cost, eviction coherence, and end-to-end
latency are not measured here (Section~\ref{sec:conclusion}).

\section{Limitations and Disclosures}\label{sec:limits}

Five limits bound the claims. First, the fingerprint uses public
parameters, so an adversary who controls content can forge collisions;
identity here is an integrity check against accidental divergence, the
caveat vLLM's documentation also states for its hash-keyed
cache~\cite{vllm2025apc}. Both implementations take the base as a
parameter, and the mitigations buy different guarantees. A secret base
drawn uniformly from $[2, p-1]$ keeps the collision probability of two
distinct strings of length at most $n$ below $n/p$, since their
fingerprint difference is a nonzero polynomial in $x$ of degree below
$n$ (the $+1$ offset keeps it nonzero across lengths): about
$1.7 \times 10^{-12}$ per comparison at four million bytes, with every
rule and bound unchanged, but only for content chosen independently of
the key and fingerprints kept private, since the fingerprint of a known
two-byte string reveals $x$. A wider or second modulus (the Python
implementation accepts $p = 2^{127}-1$) lowers only the accidental
rate. Verify-on-match is exact against any adversary: equality has no
false negatives, so a collision can only lengthen the returned prefix,
and a byte comparison of that prefix before reuse, linear in its
length, certifies it. Second, the identity layer is conditional on use
(Sections~\ref{sec:competitive} and~\ref{sec:realstack}): the
short-prefix and one-against-many regimes belong to the radix matcher,
and the value concentrates on long shared prefixes and settings outside
a single engine's cache. Third, Python multipliers carry an interpreter
constant (Section~\ref{sec:mechanisms}). Fourth, the serving wall belongs to
the engine configuration (Section~\ref{sec:realstack}), and generation
quality at extreme lengths was not scored; the engine-side claims are
timing and energy only. Fifth, the real traces replayed here are tree-of-thought search
and ShareGPT and LMSYS-Chat-1M streams, not recorded agent sessions
(Section~\ref{sec:conclusion}). A per-seed provenance gap in two of the
three synthetic base contexts is documented in the project record
(recorded corpus hashes are authoritative; results are unaffected).

\section{Related Work}\label{sec:related}

Ropes were introduced as heavyweight strings with logarithmic
concatenation and substring operations~\cite{boehm1995ropes}, and
production descendants such as Ropey~\cite{ropey} remain the editing
specialist. Weight-balanced trees~\cite{nievergelt1973bounded} and the
balanced join~\cite{adams1993efficient} supply the balancing scheme;
persistence with structural sharing follows the classical
treatment~\cite{driscoll1989persistent}. Rolling and polynomial
fingerprints originate with randomized pattern
matching~\cite{karp1987efficient}, and per-node hashes over trees with
Merkle's construction~\cite{merkle1987digital}; \hr{} carries a
Karp-Rabin-style fingerprint at every node of a rebalancing, editable
rope, which is the combination the four mechanisms need. On the
serving side, PagedAttention~\cite{kwon2023pagedattention} shares KV
blocks through paged copy-on-write tables, SGLang's
RadixAttention~\cite{zheng2024sglang} matches reusable prefixes in a
radix tree over cached tokens, and Prompt
Cache~\cite{gim2024promptcache} reuses attention states of declared
prompt modules. These systems reuse KV state given a match; \hr{}
addresses the representation question underneath them: one persistent
object that is simultaneously the editable context, the forkable
snapshot, and a verifiable content key.

\section{Conclusion}\label{sec:conclusion}

Four mechanisms of LLM context management reduce to one persistent
weight-balanced rope with per-node polynomial fingerprints. The
structure edits and answers verified identity queries in logarithmic
time at once, holds its own against each specialist on its home ground,
dominates once identity enters the workload, and on a live 7B stack
identifies a one-million-token shared prefix exactly and faster than
the production matcher, at a length the dense engine cannot serve,
while measured engine energy prices what such reuse is worth.

The applications reach past a single engine, because the structure is
domain-agnostic over bytes and the crossovers say where it pays. A fingerprint that is maintained rather than recomputed
travels: it can route requests across engine instances, deduplicate
across requests and tenants, and survive cache eviction, none of which
a per-instance radix tree provides. The same edit-plus-identity regime
appears wherever long strings are edited and compared, in versioned
document stores and collaborative editors (branch and snapshot plus
repetition encoding give logarithmic-memory histories), in
storage-layer deduplication keyed by maintained fingerprints instead
of rescans, and in multi-megabyte sequence data such as genomic
strings, whose lengths sit beyond every crossover measured here. In each case the query-rate ablation gives the
precondition: the win scales with how often the surrounding system
asks whether content is identical.

Keyed or verify-on-match hashing would close the adversarial gap the
fingerprint leaves open. A recorded-agent-trace study would replay real sessions
to measure where deployed operation mixes, context sizes, and
identity-query rates sit against the crossovers, upgrading the
motivating hypothesis of Section~\ref{sec:intro} to a tested claim. A
full serving-loop replay (per-request match and insert against a live
cache pool) extends the single-query identification study, and a
paired host-energy study of logarithmic edits against linear copies
would complete the energy story on the CPU side. Finally, rerunning
the Python-side competitive legs on the Rust implementation would shed
the interpreter constant that the guards already show is not
structural. The implementations are public at
\url{https://github.com/jemsbhai/hashrope}, and the full experimental
record (pre-registered plans, logbook, results, and figure scripts) at
\url{https://github.com/jemsbhai/hashrope-paper}, tag
\texttt{ictai2026-camera-ready}.

\bibliographystyle{IEEEtran}
\bibliography{refs}

\end{document}